# Trustworthy Autonomous Systems (TAS): Engaging TAS experts in curriculum design

Mohammad Naiseh
*School of Electronics and Computer Science*
*University of Southampton*
Southampton, United Kingdom
m.naiseh@soton.ac.uk

Caitlin Bentley
*Information School*
*The University of Sheffield*
Sheffield, United Kingdom
c.m.bentley@sheffield.ac.uk

Sarvapali D. Ramchurn
*School of Electronics and Computer Science*
*University of Southampton*
Southampton, United Kingdom
sdr1@soton.ac.uk

*Abstract*— Recent advances in artificial intelligence, specifically machine learning, contributed positively to enhancing the autonomous systems industry, along with introducing social, technical, legal and ethical challenges to make them trustworthy. Although Trustworthy Autonomous Systems (TAS) is an established and growing research direction that has been discussed in multiple disciplines, e.g., Artificial Intelligence, Human-Computer Interaction, Law, and Psychology. The impact of TAS on education curricula and required skills for future TAS engineers has rarely been discussed in the literature. This study brings together the collective insights from a number of TAS leading experts to highlight significant challenges for curriculum design and potential TAS required skills posed by the rapid emergence of TAS. Our analysis is of interest not only to the TAS education community but also to other researchers, as it offers ways to guide future research toward operationalising TAS education.

*Keywords—Trustworthy Autonomous Systems, Education, Skillset.*

## I. Introduction

Rapid advances in Artificial Intelligence (AI), including machine learning algorithms, sensors and robotics, as well as their supporting infrastructures, have enabled the growth of autonomous systems across industries and sectors [1]. Taken together, an autonomous system is one that involves software applications, machines and people, and is able to take actions with little or no human supervision [2]. Such systems promise a better and safer world where, e.g., autonomous vehicles could reduce the number of road accidents or robots could assist doctors to perform complex surgeries.

Developing autonomous systems for implementation across societies requires them to be trustworthy by design [8,9]. Trustworthy Autonomous Systems (TAS) has been defined by the UKRI TAS Hub[1] as autonomous systems whose design, engineering, and operation ensures they generate positive outcomes and mitigate potential harmful outcomes. TAS can depend on a number of factors including but not limited to: (a) explainability, accountability and understandability to a diverse set of users [4,10], (b) their robustness in dynamic and uncertain environments [11], (c) the assurance of their design and operation through verification and validation processes [12], (d) the confidence they inspire as they evolve their functionality [18], (e) their defences against attacks on the systems, users, and the environment they are deployed in [18], (f) their governance and the regulation of their design and operation, and (g) the consideration of human values and ethics in their development and use [7,18].

TAS applications and societal contexts within which they will be embedded are likely to be significantly different than they are today, for instance with robots autonomously performing tasks now carried out by humans, by humans and robots working together, or by integrating autonomous decision-making throughout our environments. Ensuring the trustworthiness of autonomous systems is a known and well-argued problem in the literature [1,2], yet there is a clear lack of attention to the skills set needed to prepare future TAS engineers and up-skill current workers to meet TAS systems requirements and values. In a survey with 118 AI organisations in the UK, many organisations reported that they went beyond AI technical skills and opened new roles that combine both AI skills and domain-specific knowledge (e.g., law, engineering) [19]. The participating organisations reported that they faced difficulties to find appropriate candidates with skills to develop the trustworthiness of autonomous systems in context. Also, in another study, AI employers in the UK reported that new graduates lacked practical experience and understanding of AI or its potential impact on societies and individuals [20]. This implies that graduates are not equipped to apply their training to real-world situations and that engineering education needs to change.

These limited findings imply that TAS practitioners require different skill sets and different training than what they receive. These skills may include engineering the robustness, reliability of TAS, but also in how human values and ethics are enacted, as well as the explainability, accountability and understandability of TAS to diverse sets of actors. To that end, the UKRI Trustworthy Autonomous Systems Hub (TAS-Hub) has launched a Syllabus Lab, which is a collaborative project between academic institutions, industry partners and professional organisations to facilitate upskilling and reskilling within key industries. Syllabus Lab aims to better understand how social, technological, legal and ethical values of TAS should be translated into educational approaches and materials, effectively preparing the future generation of TAS engineers as well as up-skilling current ones. A second aim of the Syllabus Lab is to develop innovative ways to share resources and structured courses across universities and schools to scale the impact of this work. We will also likely need to incorporate and share tools and techniques from across disciplines, as well as define a skill set that can respond to emerging TAS requirements.

---

[1] The Hub sits at the centre of the Trustworthy Autonomous Systems Programme, funded by the UKRI Strategic Priorities Fund. https://www.tas.ac.uk/

As a first step towards meeting these aims, this study brings together the collective insights from the workshop entitled "TAS Skills Development Workshop", held by members of the Syllabus Lab on 14th September 2021. Contributions were received from collaborators within industry, academia and the public sector to highlight the significant challenges and potential required skills for future TAS engineers.

## II. RELATED WORK

Trustworthiness is mainly a socio-cultural concept [21], meaning that Trustworthy Autonomous Systems necessarily involve understanding and addressing both socio-cultural and technical factors together in context. Tackling the TAS curriculum from a purely technical view runs the risk of making abstraction errors [22], meaning that the curriculum is missing the focus on understanding the broader context in which TAS are embedded. Additionally, the development and deployment of TAS typically involve diverse, uncertain and complex decisions that require making trade-offs during the TAS development process [17]. The challenge lies in that the performance and the quality of TAS must align with different properties and values of TAS. For instance, trade-offs can happen in TAS design when engineers need to decide how to balance explainability or usability properties. In this case, TAS practitioners could introduce friction to the user's experience to encourage them to pay attention to a robot's explanation of its actions to promote safe and effective Human-Robot collaboration. Therefore, consistent with the literature on designing systems with socio-cultural principles in mind, e.g., [23,24], we need empirically grounded research on the TAS skills and knowledge needed for engineers to engage with this complexity.

Concerning the current state of research in this area, a variety of public- and private-sector organisations have published principles, skills frameworks, or curriculum design recommendations intended to guide autonomous systems education in universities and schools, as well as bridge the gap in the industrial workplace. High-profile examples include the UK-RAS (Robotics & Autonomous Systems) Network [13], Deloitte - Essential skills for humans working in the machine age [15], IKE – Innovation Skills Development Framework [14], and UK Government DDT (Digital, Data and Technology) framework [16]. However, these frameworks have been constructed for domain-specific applications. In terms of their content, one focuses too much on technical skills [13], another is too domain-centric [25], while yet another is too broad and presents general skills needed across different jobs in the machine era [15]. Despite the relevance of prior frameworks, none of them proposes a conceptual model representing the profile of a TAS engineer in an organisation, or gives a methodology describing the TAS knowledge base and skills set based on both academic and industrial views. As a result, current frameworks can fail to achieve intended TAS goals if they are not accompanied by other values and multidisciplinary approaches for ensuring that practitioners discover all facets of TAS.

## III. WORKSHOP DESIGN AND WORKFLOW

As discussed above, many organisations have published autonomous systems skills frameworks; however, their domain-specific nature makes them difficult to operationalise. As an initial step to address this gap, we conducted a workshop with TAS experts, whom we invited during TAS-Hub All Hands Meeting (AHM)[2]. Researching the values and considerations of TAS experts seems a crucial first step for both designing a TAS curriculum, and understanding where there are significant gaps to address. The workshop brought together the TAS-Hub community to talk about TAS education, think about main challenges, discuss TAS skills, and steer the overarching strategy for the coming year. The workshop lasted for three hours, spanning across two main sessions. Session 1 consisted of an introduction and "TAS skills" presentations to familiarise and engage TAS experts in the research problem (See Table I). Session 2 consisted of a co-design activity and discussion (See Table II). Next, we describe the workshop design and the activities in detail.

*Session 1.* This part of the workshop started with welcome and introduction presentations. Experts were also given a Miro link [6] – the collaborative whiteboard tool used throughout Session 2. Then, the research team delivered an overview presentation for the current skills-related framework in the literature and discussed main gaps and limitations. After that, three different presentations were delivered to present best practices in TAS education from real-world case studies – (a) preparing the workforce in the health sector for AI and Autonomous Systems (NHS AI Lab), (b) TAS skills: an industrial perspective (Thales), and (c) The 3A Institute's approach to course co-design with students and industry partners (Australian National University). Each presentation lasted for 15 mins. Finally, Session 1 ended with a discussion and invitation for TAS experts who would like to contribute to TAS skills discussion and design (Session 2). A total of 40 TAS experts participated in Session 1.

TABLE I. SCHEDULE FOR SESSION 1

| Time | Activity |
| --- | --- |
| 15 min | Welcome and Introduction |
| 15 min | Overview of AI-related skills frameworks, strategies and gaps |
| 45 mins | Exploring best practices in trustworthy autonomous systems education |
| 15 min | Discussion and Q&A |

---

[2] The All Hands Meeting (AHM) brought together TAS community to talk about their research and steer the overarching strategy for the coming year. https://www.tas.ac.uk/bigeventscpt/all-hands-meeting/

TABLE II. SCHEDULE FOR SESSION 2

| Time | Activity |
|---|---|
| 60 min | Co-design for TAS skills development (Three parallel sessions)<br>1. TAS capability and skills<br>2. TAS curriculum design (Academic perspective)<br>3. TAS curriculum design (Industry perspective) |
| 15 min | Session feedback and discussion for future work |

*Session 2.* Fourteen TAS experts (or 35% from Session 1) joined the second session who were able to commit to the workshop time and had an interest in providing a contribution based on their experience. However, we are concerned that more TAS experts chose not to stay for the second session. This could have been due to interest in other parallel sessions offered at the workshop. In the future, we would like to investigate the reasons why so many experts choose not to engage actively in TAS educational research. Experts were assigned to three different groups based on their experience and TAS background. We chose to split experts into three groups to capture different perspectives on TAS education. Each group discussed 3-4 questions to reflect on the current state of TAS and discuss what is needed for the future. The three groups are:

*1) TAS Capability and Skills Frameworks:* this group discussed key skills needed for future TAS engineers.

*2) TAS university courses:* this group discussed what is needed in universities to prepare future TAS engineers.

*3) TAS industry training:* this group discussed industry needs for future TAS engineers.

The questions across the groups broadly focused on:

*1)* Key issues or challenges in TAS education from each perspective.

*2)* Skills, approaches, or initiatives needed to address the key issues or challenges, or any existing best practice examples.

*3)* Key recommendations for the TAS Hub Syllabus Lab moving forward.

Our grouping rationale was to ensure the diversity and credibility of the collected data as well as capture the diverse interest and backgrounds of our TAS experts. Each group was paired with at least one researcher, who led the discussion and answered participants questions. We used Miro [6] to ask participants to answer the above questions and to collectively decide on the top three issues we should prioritise. Each group was successful in producing a rich representation of their discussion on the Miro board, as well as in deciding on the top three priorities. A screenshot of the Miro board can be found in Figure 1, which is not shown to scale in order to maintain the confidentiality of our experts.

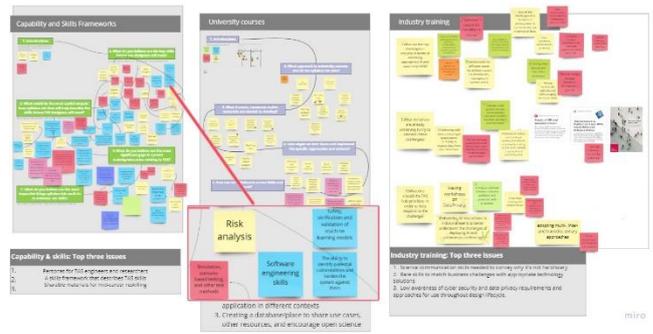

FIGURE 1 CO-DESIGN ACTIVITY

IV. FINDINGS AND IMPLICATIONS

The workshop prompted a rich discussion with three groups of experts completing three multidisciplinary discussion views. The views were: (1) TAS capability and skills frameworks, (2) TAS university courses, and (3) TAS industry training. Our analysis is a result of TAS expert discussions and co-design activity within a diverse group of researchers, educators, and industry practitioners. While not comprehensive, we believe our analysis captures a broad range of opinions from multiple stakeholders and synthesises a feasible way forward. In the following sections, we first present draft TAS skills set that emerged from our workshop. We then present the main recommendations to address in future research.

*TAS capability and skills frameworks.* TAS experts identified a range of skills and related attributes relevant to building TAS. Essential elements of the skills are presented in Table III. It is important to note that the TAS profession will include a variety of job roles in the future and these might require varying levels of skills. For example, a machine learning (ML) engineer will be more involved with writing code or evaluating and verifying TAS during the development process. On the other hand, a user experience expert will be more involved in creating and developing usable interfaces for end-users. Irrespective of the specific role, all TAS professionals are likely needed to have proficiency in the three pillars of soft, TAS technical and strategic skills. The following sections present three groups of skills that emerged in our workshop and discuss their relevance to TAS.

- *Soft skills* are important to enable successful TAS development. Soft skills have long since been shown to be a crucial component of engineering education, specifically, a number of studies point to the ineffective communication skills of information systems staff as a possible cause of failed projects [3,5]. However, our experts discussed how TAS contexts are unique in that TAS professionals are dependent on interacting with people across disciplines and professions across different stages of TAS development processes. TAS professionals in an organisation cannot work on an island but are required to work closely with end-users and different actors involved in the project, or alternatively, develop a keen sense of interpersonal dynamics in cross-cultural societies and diverse regulatory environments. This requires good communication, collaboration and agile management skills. For instance, TAS engineers involved in implementation processes need to understand emergent trust issues and respond to these issues sensitively.

- *TAS technical skills.* Technology plays an integral role in TAS and requires professionals to possess a diverse set of technical skills ranging from programming, AI concepts and algorithms, data analysis, risk assessment, human factors testing and engineering, software engineering, hardware design, explainability design and testing, AI ethical decision-making, and bias detection and mitigation. All the skills listed in Table III were accepted by our TAS experts as necessary for all the roles that TAS engineers will have in the future. They argued that it is also important to have a basic understanding of the core technical areas in order to have a holistic view of TAS.

- *Strategic skills* help to enable and initiate strategical thinking about TAS projects, including planning for future improvements and challenges. In general, our experts discussed that TAS professionals are responsible for making trust-related design decisions which require being aware of the changing trends in the field and leveraging innovative ideas that will help successful implementation. These responsibilities will require strategic skills which include, leadership, self-learning, adaptability to change, understanding of regulation and legality of TAS, creative and critical thinking.

TABLE III.     DRAFT TAS SKILL SET

| Dimension | Skills |
|---|---|
| Soft | Communication<br>Management<br>Teamwork |
| Strategic | Self-management<br>Self-learning<br>Problem-solving<br>Creative thinking<br>Critical thinking<br>Adaptability and flexibility<br>Understanding regulation and legality of TAS |
| TAS technical | TAS ethical decision-making<br>Human factors<br>Software engineering<br>User experience<br>Risk assessment<br>Explainability design and testing<br>Bias detection and mitigation<br>Hardware design<br>Machine learning<br>Programming |

*TAS recommendations and future directions.* During the workshop, all groups discussed current education gaps and made recommendations on future research directions to prepare and up-skill TAS engineers. Our TAS experts across three groups identified four main recommendations they valued to be important for TAS education: *increase the interdisciplinary awareness in TAS educational courses, prepare appropriate logistics to successfully deliver TAS courses, increase diversity in the TAS workforce and identify skills required for different TAS roles.*

*Increase the interdisciplinary awareness in TAS educational courses.* Experts discussed that TAS is a highly interdisciplinary and dynamic field of study and practice, comprising multifarious research that is scattered across disciplines including psychology, sociology, economics, management, law, computer science, and so on. During the co-design sessions, all groups discussed that promoting the awareness of the interdisciplinary nature of TAS systems is the main gap in current TAS education and it is critical to developing tolerance and sensitivity to differences in underlying epistemological, ontological or pragmatic orientations within a curriculum. Our experts frequently added notes referring to awareness of the nature of TAS, suggesting that the TAS curriculum should adopt multi-, inter- and transdisciplinary courses and tools. Experts further discussed that TAS education should also promote interdisciplinary awareness following a two-way relationship model involving: (1) the impact of autonomous systems on other disciplines; and (2) how other disciplines can also affect the design of autonomous systems. For instance, experts discussed that TAS courses could be supported with examples of successful projects and cases studies following a two-way relationship model — in the form of papers, books or other media effectively co-authored by scholars of different disciplinary perspectives. Experts also suggested encouraging or allowing students to tangibly collaborate across disciplines as part of their required assessment for coursework. Multi-disciplinary collaboration can be made easier by running concurrent or joint courses with other departments, so that students may gain practical project experience engaging with collaborators of a different disciplinary lens.

*Prepare appropriate logistics to successfully deliver TAS courses.* Logistics refers to the factors that enable the curriculum to be successfully delivered in practice. Experts discussed that the limited availability of TAS resources, specifically human resources, which makes TAS education challenging to deliver due to the interdisciplinary nature of the subject matter. TAS education must clearly cover key concepts ranging from engineering to ethics and psychology as well as the use of those concepts to analyse applications of autonomous systems. Moreover, TAS courses may also cover computational approaches to trust; psychological or sociological theories of trustworthiness; and, methods for experimentally, participatively or interpretively investigating trust issues. This requires lecturers with broad interdisciplinary backgrounds and constructivist pedagogical methods to draw on knowledge from disparate disciplines, or, the capacity to collaborate to teach in multidisciplinary teams effectively. Design and delivery of new TAS university courses is also complicated, as multiple departments are likely needed to cooperate. Furthermore, experts saw the multidisciplinary nature of TAS as a challenge that is still under investigation, finding it difficult to imagine core topics and stills before more research is done. TAS experts highlighted various logistics issues that should be addressed in TAS education to successfully implement TAS curriculum. These logistics included, partnerships with leading industries to provide internships, appropriate labs, simulation tools and databases for successful TAS projects. These suggestions highlight the situated nature and tight coupling of TAS with industry and practice settings.

*Increase diversity in the TAS workforce.* Experts also discussed that TAS education should have a diversity policy to increase the representation of women and ethnic minority groups that are marginalised in the AI sector. This is particularly important for TAS to mitigate the risk that a non-diverse workforce could perpetuate or accelerate bias; a diverse group of TAS engineers reduces the risk of the inherent bias embedding into TAS designs. Making the TAS workforce more diverse will also be important in order to ensure that TAS works well for a wide range of different users. Experts also added that TAS should avoid the highly remunerated and rapidly growing career opportunities being dominated by already privileged groups in society, thereby increasing inequality. An implication of this for TAS education is that universities and industry partners alike collaborate on defining clear diversity and inclusion targets, as well as sharing transparently their progress on these matters.

*Identify skills required for different TAS roles.* It was also suggested that the focus should move away from thinking about primarily about increasing the number of TAS engineers, towards thinking about TAS skills required for different (and often already existing) roles, such as project managers, executives, software engineers, or user experience designers to name a few. Our experts felt it would be helpful to focus on the core skills required across different TAS roles and how educational material could be adapted or personalised to ensure students gain the specific skills needed for diverse TAS roles. We believe further research is needed to ascertain how important it is to embed education to develop the broad TAS skillset (i.e. outlined in Table III) for all.

## V. Conclusion and future work

In conclusion, Trustworthy Autonomous Systems is an emergent area discussed in academia and education. There is a serious lack of relevant studies, particularly in planning, implementing and conceptualising TAS curricula. The unique challenge about TAS is that it is new, emerging, interfering and disruptive. It thus offers an opportunity to open the discussion and identify boundaries for identifying key skills required for future TAS engineers. In this paper, we identified three sets of skills that reflect an interdisciplinary approach that focuses on soft, strategic and TAS technical skills. The study concludes by identifying the various recommendations that future TAS education shall consider to promote an effective interdisciplinary approach to TAS education. Our next work will involve a host of funded research projects in the TAS Hub network that will come to a stage that will enable the translation of research outcomes into educational materials [12]. From these projects, we will develop use cases and case studies to publish open access. Our subsequent research will focus on the intermediation mechanisms required to support the take-up and integration of such open educational resources across the Hub's network of university and industry partners. Through this experience we will carry out mixed methods research to monitor and understand the impact of these resources on the aspects outlined in this paper: 1) how the resources support TAS skills development, and/or whether these are the 'right' skills to emphasise at this stage; 2) the logistics needed to implement and use the resources effectively, and 3) how the resources support diversity and inclusion principles in TAS education. We invite and welcome any partners and collaborators in the engineering education community to join us in this venture.

## Acknowledgement

This work was developed from a workshop "TAS Skills Development Workshop", held by members of the Syllabus Lab on 14th September 2021. We are thankful to our TAS experts who attended the workshop and contributed their perspectives. This work was conducted as part of the Trustworthy Autonomous Systems Hub, supported by the Engineering and Physical Sciences Research Council (EP/V00784X/1).